\documentclass[12pt,preprint]{aastex}

\def\ifundefined#1{\expandafter\ifx\csname#1\endcsname\relax}

\usepackage{lscape}

\def\la{\mathrel{\hbox{\rlap{\hbox{\lower4pt\hbox{$\sim$}}}\hbox{$<$}}}}
\def\ga{\mathrel{\hbox{\rlap{\hbox{\lower4pt\hbox{$\sim$}}}\hbox{$>$}}}}

\newcommand{\be}{\begin{eqnarray}}
\newcommand{\ee}{\end{eqnarray}}

\ifundefined{ensuremath}\def\ensuremath#1{\relax\ifmmode{#1}}
\else${#1}$\fi\else\relax\fi
\ifundefined{nuc}\def\nuc#1#2{\relax\ifmmode{}^{#1}{\protect\textrm{#2}}
\else${}^{#1}$#2\fi}\else\relax\fi

\newcommand{\kmps}{\ensuremath{\mathrm{km}~\mathrm{s}^{-1}}}

\newcommand{\zsol}{\ensuremath{{\mathrm{Z}_\odot}}}

\def\Tmod{\ensuremath{T_{\mathrm{model}}}}

\def\tstd{\ensuremath{\tau_{\mathrm{std}}}}

\newcommand{\vno}{\ensuremath{v_0}}

\newcommand{\phx}{\texttt{PHOENIX}}

\newcommand{\gamray}{$\gamma$-ray}
\newcommand{\halpha}{H$\alpha$}

\shortauthors{Ketchum, W. et~al.}
\shorttitle{Detailed Spectral Analysis of SN 1999dn}

\begin{document}

\title{Detailed Spectral Analysis of the Type Ib Supernova 1999dn.
Paper I:  Hydrogen-free Models}

\author{ 
  Wesley Ketchum,
  E.~Baron\altaffilmark{1},
  and David Branch
 }
\affil{Homer L.~Dodge Department of Physics and Astronomy,
   University of Oklahoma, 440 West Brooks, Rm.~100, Norman, OK
   73019-2061 USA}
\email{ketchum,baron,branch@nhn.ou.edu}

 \altaffiltext{1}{Computational Research Division, Lawrence Berkeley
   National Laboratory, MS 
   50F-1650, 1 Cyclotron Rd, Berkeley, CA 94720-8139 USA}

\begin{abstract}
We present spectral fits to five epochs of the  typical Type Ib
supernova 1999dn using the generalized, non-LTE, stellar atmospheres
code \phx.  Our goal is threefold:  to determine basic physical
properties of the supernova ejecta, such as velocity, temperature, and
density gradients; to reproduce He I absorption lines by invoking
non-thermal excitation; and, to investigate possible spectral
signatures of hydrogen, especially a feature around 6200 \AA, which
has 
been attributed to high velocity \halpha.  Our models assume an
atmosphere with uniform composition devoid of any hydrogen.  Our
model spectra fit the observed spectra well, successfully reproducing most of
the features, including the prominent He I
absorptions.  The most plausible alternative to \halpha~as the source
of the 6200 \AA~feature is a blend of Fe~II and Si~II lines, which can
be made stronger to
fit the observed feature better by increasing the metallicity of the ejecta.
High-metallicity models fit well at early epochs, but not as well as
solar-metallicity models after maximum light.  While this blend of
metal lines is a reasonable explanation of the source of the 6200
\AA~feature, it is still important to investigate hydrogen as the
source; therefore, a second paper will present models that include a
thin shell of hydrogen around the main composition structure.
\end{abstract}

\keywords{stars:  atmospheres---supernova:  individual (SN 1999dn)}

\section{Introduction} 

Supernovae (SNe) are divided into two different types, I and II, based
on the presence of hydrogen in their optical spectra.  Type II
supernovae spectra have conspicuous hydrogen lines while Type I
supernovae spectra do not.  SNe~I can further be divided into three
more spectral types:  Ia, Ib, and Ic.  SNe~Ia characteristically
contain strong absorption lines caused by singly-ionized silicon;
SNe~Ib contain strong neutral helium absorption lines; and, SNe~Ic
lack both of these characteristics \citep{filarev97}.

Because of their almost exclusive location in the star-forming regions
of spiral galaxies, and because of their spectral properties, it is
believed that Type II, Ib, and Ic supernovae all result from the
core-collapse of massive stars.  Furthermore, studies have shown that
SNe Ib/c are likely to result from binary star systems \citep{heger03}.

Until recently, the very small number of SNe~Ib spectra available near
the period of maximum luminosity limited the study of these events in
detail.  However, a great number of Ib spectra have become available
thanks to the published spectra of \citet{math_sn1bc} and others.
Recent work with these spectra has shown that there may be traces of
hydrogen in SNe~Ib spectra, and of both hydrogen and helium in SNe~Ic
\citep{branIb02,branch_hIc06,elm_1bc06,parrent05bf07}.  This work used
SYNOW---a highly-parameterized, supernova spectrum-synthesis code that
assumes a sharp photosphere that emits a blackbody continuous spectrum
and line formation by resonance scattering, treated in the Sobolev
approximation---to calculate spectral models.  Other work on modeling
the spectra of SNe~Ib has also used SYNOW
\citep{deng00,benetal91D02,anupama05bf05}.

The speed and relative simplicity of SYNOW makes it very useful for
identifying spectral lines, line velocities, and other constraints on 
the ejected mater \citep{branIb02}.  However, these model spectra
do not naturally account for non-thermal excitation, which is required to see
He I absorption lines; the optical depth of He I absorptions is an adjustable
modeling parameter.  For this purpose, as well as to gain more
information about the physical properties of SNe Ib, we conduct a more detailed
spectral analysis using the generalized, non-LTE, stellar atmospheres
code \phx\ \citep{hbjcam99} to complement this work.  We choose to study SN~Ib
1999dn:  a typical, well-covered Ib supernova
\citep{deng00, math_sn1bc}.  SN 1999dn has also been modeled
using SYNOW by \citet{deng00} and \citet{branIb02}.  It is
believed by both sets of authors that an absorption feature
around 6200 \AA~is most likely \halpha at early times, though they
leave open as possibilities Ne~I, Si~II, and C~II.  \citet{deng00}
argue that the feature later evolves into C~II in the post-maximum
light spectra; \citet{branIb02} suggest that the feature may be
attributed to lines of Fe~II.

We hope to accomplish the following in a two-paper study:  (1) determine
basic properties of the supernova such as effective temperature and
velocity of the photosphere, (2) determine if He~I absorption lines
can be properly formed by including non-thermal excitation, and (3)
study the issue of the existence of spectral signatures of hydrogen,
particularly determine the most plausible source of the 6200
\AA~feature.  Here we present synthetic spectra of SN
1999dn assuming that no hydrogen is present; we use a homogeneous
helium core as a basic model atmosphere.  \S 2 contains an
overview of the supernova and the observed spectra, \phx\ code
calculations, and the basic parameters adjusted to better our fits.
\S 3 features our best fits for five of the observed spectra with a brief
analysis of each fit.  Also, we investigate some of the alternatives
to \halpha~for the 6200 \AA~feature.  In \S 4 we offer a discussion of
our results and a look forward to Paper II where we will add hydrogen
to the models.

\section{Methods}

\begin{figure}
\leavevmode
\begin{center}
\includegraphics[width=0.75\textwidth,angle=90]{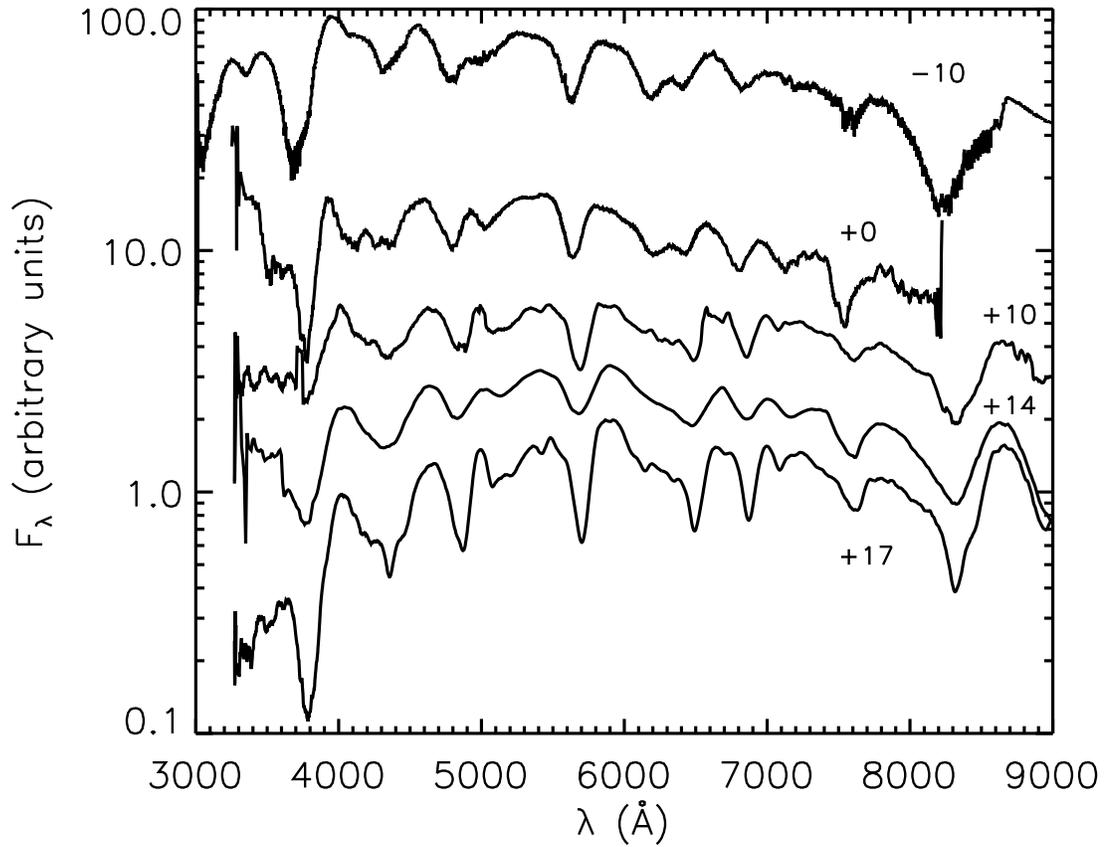}
\end{center}
\caption{\label{fig:all}The five observed spectra we model, referenced
  by their epochs, plotted in semi-log with arbitrary shifts added to
  make the spectra easier to see.  The -10, +0, and +14 day spectra
  are from \citet{deng00}.  The +10 and +17 day spectra are from
  \citet{math_sn1bc}.}
\end{figure}

SN~1999dn was discovered on August 20 by \citet{qiu99} in NGC 7714,
which has a recession velocity of 2700 \kmps~and an $E(B - V) = 0.052$
\citep{schlegelred98}.  In total, six spectra of SN 1999dn
have been published, ranging from ten days before until 38 days after
maximum brightness in the R-band, which occurred on August 31
\citep{math_sn1bc}.  The five spectra we model are shown in 
Figure~\ref{fig:all}.  While there was some early confusion about
its classification, being identified early as a Ia or Ic, SN~1999dn
contains conspicuous He I absorption lines, and so is classified as a
Type Ib \citep{math_sn1bc}.  \citet{branIb02} consider SN
1999dn a typical SN~Ib, and so a study of its spectra should provide
insight into SNe~Ib as a whole.

We present spectral fits of the first five epochs of SN 1999dn
obtained from the multi-purpose stellar atmospheres program \phx\
\citep{hbjcam99,bhpar298,hbapara97,phhnovetal97,phhnovfe296}.  \phx\ solves  
the radiative transfer equation along characteristic rays  in
spherical symmetry, taking into account relativistic effects.  The
non-LTE (NLTE) rate equations for selected ionization states
are solved, thereby including effects of non-thermal
electrons from the $\gamma$-rays produced by the radioactive $\beta$ decay
of Ni-56, which are essential to making He I and any Ne I absorptions.

Table~\ref{tab:ions} lists the 27 ions that are treated in NLTE in our
models.  Each model atom includes primary NLTE transitions, used to calculate
the level populations and opacity, and weaker LTE transitions, which
are included in the opacity, but also which affect the rate equations
because of their effect on the solution to the transport equation
\citep{hbjcam99}.  All LTE line opacities not treated in NLTE
are treated with an equivalent two-level atom source function, using
a thermalization parameter, $\alpha$ = 0.05.  The atmospheres are
iterated to energy balance in the co-moving frame; while we neglect
explicit effects of time dependence on the transport equation, we
implicitly include them by including the $\gamma$-ray deposition rate in
the radiative equilibrium equation and in the rate equations for the
NLTE populations.

All of the models shown have a uniform composition structure with
$\gamma$-ray deposition that follows the density of the ejecta to
provide for the non-thermal excitation necessary to produce the
observed He I absorption lines (and any Ne I absorption as well).  The
``standard composition'' models have solar metallicities, except
we assume all hydrogen has been burned to helium; the
``high-metallicity'' models have three times solar metallicity.  The models are
parameterized by a few quantities:  the time since explosion (with an
assumed 12-day rise time, $t$), the effective temperature (\Tmod), the
photospheric velocity (the velocity, \vno, where the continuum optical depth
in extinction at 5000 \AA~, \tstd, is unity), the metallicity (given
as a factor $Z$ times solar), and the density profile's power law
index $n$, which follows the relation $\rho \propto r^{-n}$.

\section{Spectral Fits}

We produced spectra from the calculations of over 100 models, and
compared them to the observed spectra.  Table~\ref{tab:models} contains the
adjustable parameters for the best fits for all five epochs we modeled.  The
goodness of the fit is determined in a ``$\chi$-by-eye'' fashion; the
human eye is very adept at pattern recognition.  In general, we fit
the basic shape of the continuum by varying \Tmod, fit the Doppler shift
of the spectra by varying \vno~(we most often fit the 5876 \AA~He I
absorption), and fit the widths of the observed absorption lines by
varying $n$ and \vno.

\begin{figure}
\leavevmode
\begin{center}
\includegraphics[width=0.75\textwidth,angle=90]{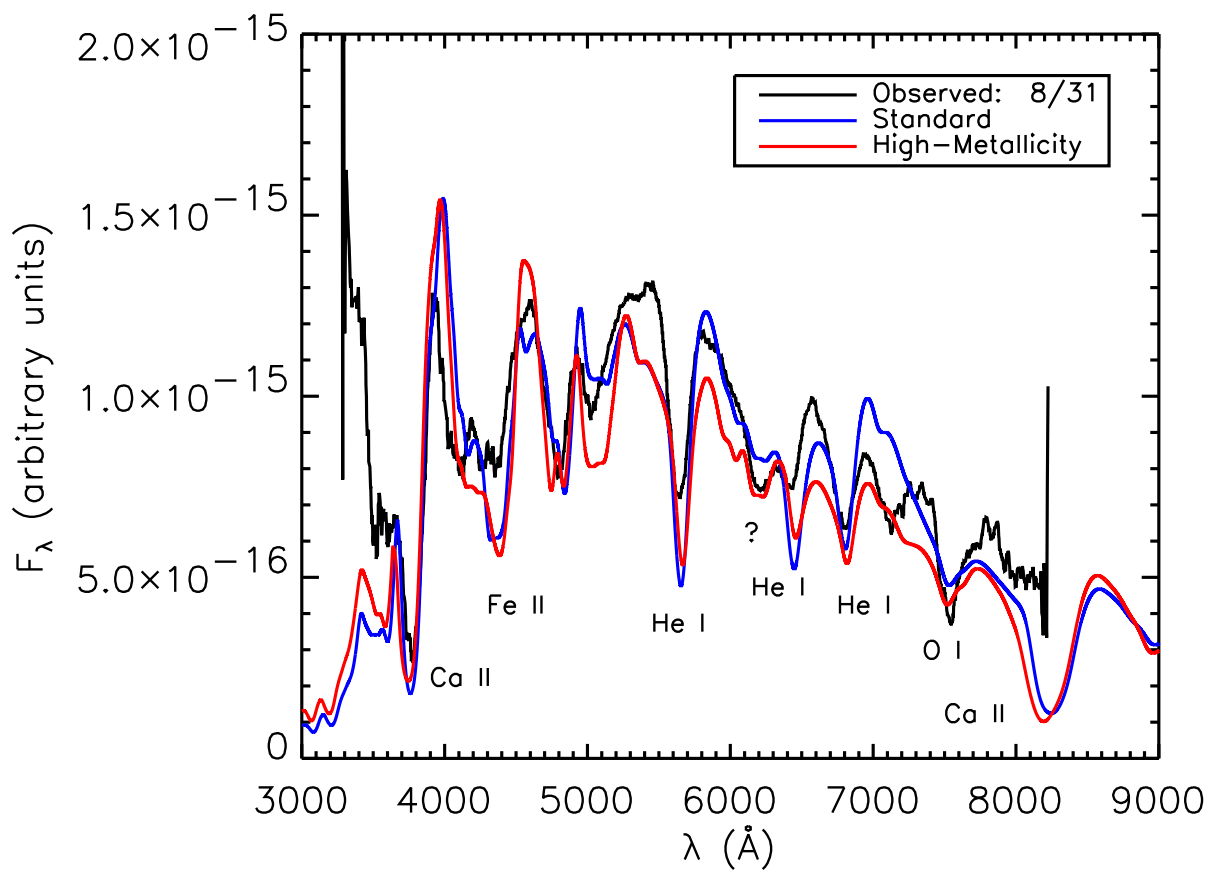}
\end{center}
\caption{\label{fig:aug31_both}The observed spectrum of SN~1999dn obtained
  on August 31, the day of maximum light (epoch +0) is compared to two PHOENIX
  models:  the first with the ``Standard Composition,'' the second
  with 3 times solar metallicity.  Both models fit the observed well,
  reproducing the He I absorptions fairly accurately.  The
  high-metallicity model fits the 6200 \AA~feature, believed to be
  \halpha,  better than the standard composition model.  In this and
  subsequent plots the observed spectra have been dereddened with $E(B
  - V) = 0.052$ \citep{cardelli89} and deredshifted using a recession
  velocity of 2700~\kmps.}
\end{figure}

Figure~\ref{fig:aug31_both} shows the day of maximum light
(epoch +0, obtained August 31) observed spectrum
and two model spectra (with parameters given in
Table~\ref{tab:models}).  For this
particular epoch we calculated 44 different models varying the basic
fitting parameters mentioned in the paragraph above, but mostly
varying the abundances of elements included in the calculations.
Identifications for the major features are shown in the figure,
corresponding to (reading the labels from left to right):  Ca~II H and
K, a multitude of Fe~II lines (with a small contribution from Mg~II,
most likely 4481 \AA;  overall it is difficult to uniquely
identify any one of the features due to the large amount of blending
in the spectra), He~I 5876 \AA, the unknown 6200 \AA~feature, He~I
6678 \AA, He~I 7065 \AA, O~I 7772 \AA, and the Ca~II
infrared triplet.  While not shown, the He~I 10830 \AA~
absorption feature is very strong in the model spectra.

Overall, both models, the standard composition and high-metallicity,
fit the observed spectrum well.  The Ca~II H and K absorptions match
the observed spectrum very well, while the emission is slightly
overshot.  The (mostly) Fe~II absorption region is on the whole fit well
by both models, though perhaps slightly better by the standard
composition model as it has more subdued absorptions and emissions.
Neither model reproduces an absorption feature near 5100 \AA~in
the observed spectrum, presumably due to Fe~II, very well.  Both
models have an absorption too deep for the He~I 5876 \AA~
feature, though the standard composition has a well matched emission.
Both models show a feature that fits the 6200 \AA~feature, with
the higher-metallicity model getting the strength nearly perfect.
Both models then produce too strong an absorption for He~I 6678
\AA~and He~I 7065 \AA, and have trouble reproducing the O~I 7772 \AA~
feature well.  Unfortunately the range of the
observed spectrum does not extend far enough to include the Ca~II
infrared triplet.

\begin{figure}
\leavevmode
\begin{center}
\includegraphics[width=0.75\textwidth,angle=90]{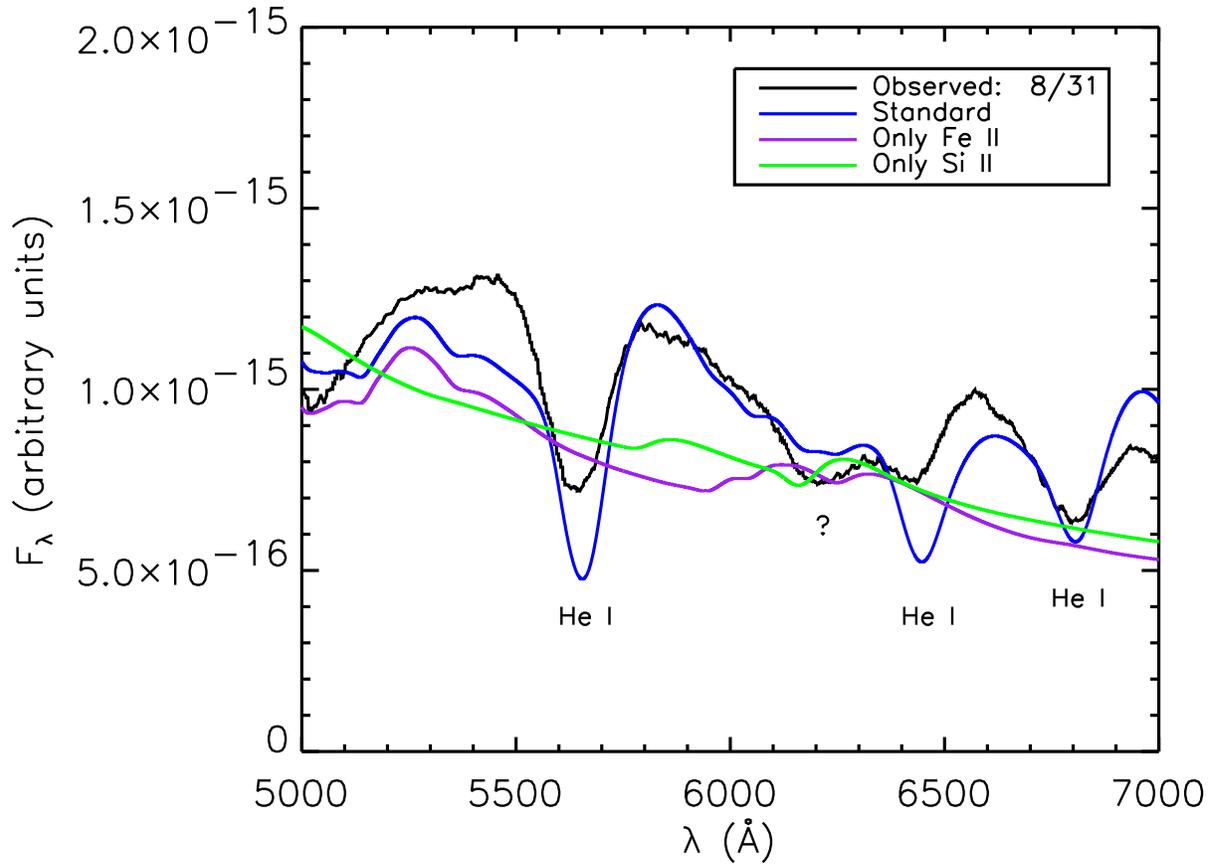}
\end{center}
\caption{\label{fig:aug31_Fe2Si2}The 6200 \AA~feature in the
  observed spectrum is compared to the feature in the standard
  composition model spectrum.  Also shown are single-ion spectra from
  the standard composition model for Fe~II and Si~II.  We can see that
  the feature in the model spectrum is a blend of these two ions.}
\end{figure}

As we can see in Figure~\ref{fig:aug31_both}, the simple non-thermal
excitation provided
in the models was enough to reproduce the neutral helium absorptions
fairly well.  Of particular
interest in this fit, however, is the good fit for the 6200 \AA~
feature, especially in the high-metallicity model.  We can see the
source of this feature in the model spectra in
Figure~\ref{fig:aug31_Fe2Si2}, which
contains the observed spectrum and standard composition model along
with single-ion spectra of Fe~II and Si~II produced from the standard
composition model.  The single-ion spectra show 
the spectral features due only to Fe~II or Si~II in the standard
composition model spectrum.  The 6200 \AA~feature in the model spectra is
the result of blending of Si II $\lambda$6355 \AA, which is formed a
little far to the blue, and an Fe II feature 
that is formed a little far to the red.  Together, they make the
feature seen in the standard composition model spectrum; their
strength is amplified by the increased abundance of both ions in the
high-metallicity model.

\begin{figure}
\leavevmode
\begin{center}
\includegraphics[width=0.75\textwidth,angle=90]{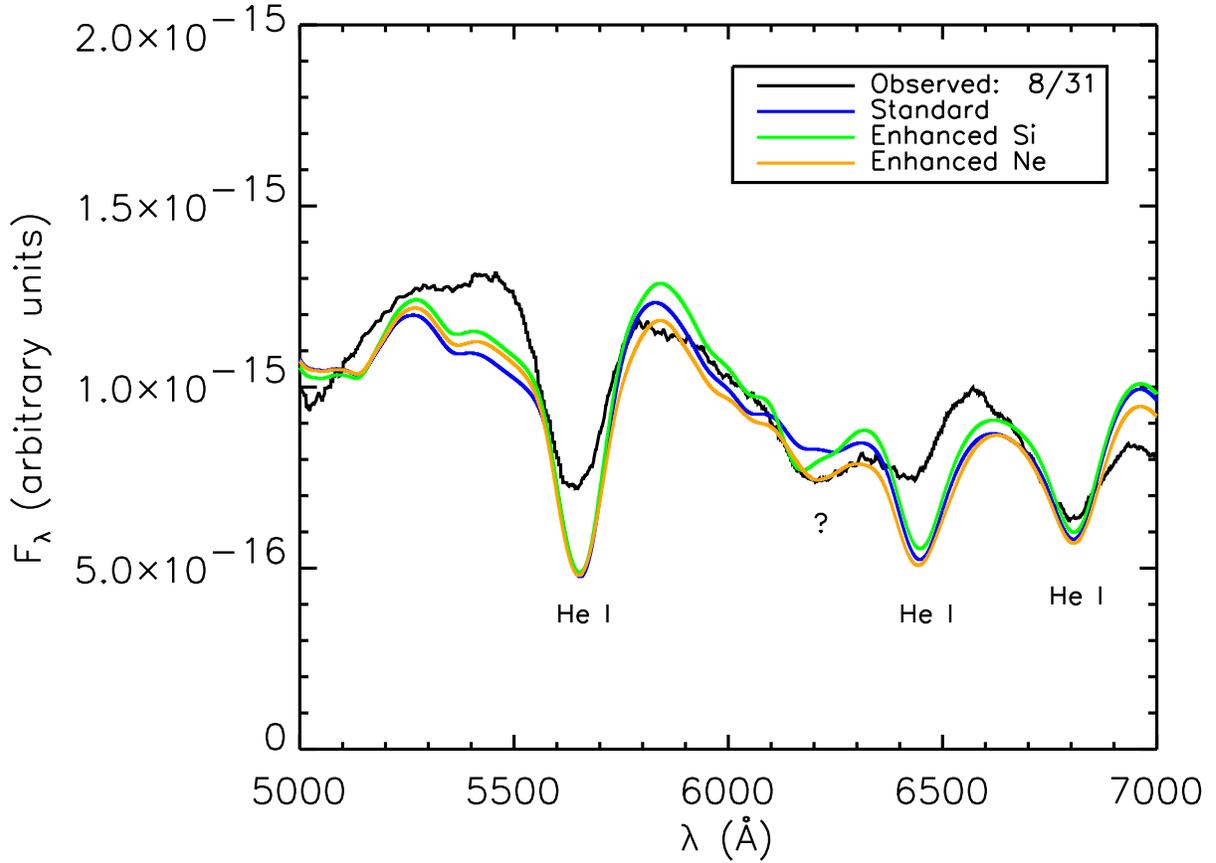}
\end{center}
\caption{\label{fig:aug31_alts}The 6200 \AA~ feature in the
  observed spectrum is compared to the feature in the standard
  composition model spectrum and two alternative models:  the first has
  10 times the Si abundance as the standard composition model; the
  second has 100 times the Ne abundance.  The increased Si abundance
  cannot account for the 6200 \AA~feature itself.  The increased Ne
  abundance reproduces the feature remarkably well, but such a
  dramatic increase in neon abundance does not seem reasonable.}
\end{figure}

Figure~\ref{fig:aug31_alts} shows the observed spectrum from the +0
epoch with the
standard composition models and two other revised models:  one with
10 times as much silicon abundance and one with 100 times as much
neon abundance.  We can see that the former revised model produces a
feature on the blue edge of the observed 6200 \AA~feature, but it does not
extend far enough to the red to be a real candidate for the source of
the absorption.  However, the latter revised model, with 100 times as much
neon as the standard composition, reproduces the 6200 \AA~feature very well.
The Ne I (6402 \AA) absorption line, produced like neutral
helium lines by non-thermal excitation, fits the feature remarkably
well, without introducing any unwanted features in other regions of
the spectrum.  Yet, such a large neon abundance does not seem reasonable.  This
could be an indication that the density profile for non-thermal
excitation is not as simple as we have modeled it, or it may be
coincidence.  While a model with enhanced carbon is not shown,
C II makes no obvious effects in the
standard composition model at this epoch, and so we disregard it as
the source of the 6200 \AA~feature as well.  The most plausible
identification from our models remains that of Si II and Fe II
blended together.

\begin{figure}
\leavevmode
\begin{center}
\includegraphics[width=0.75\textwidth,angle=90]{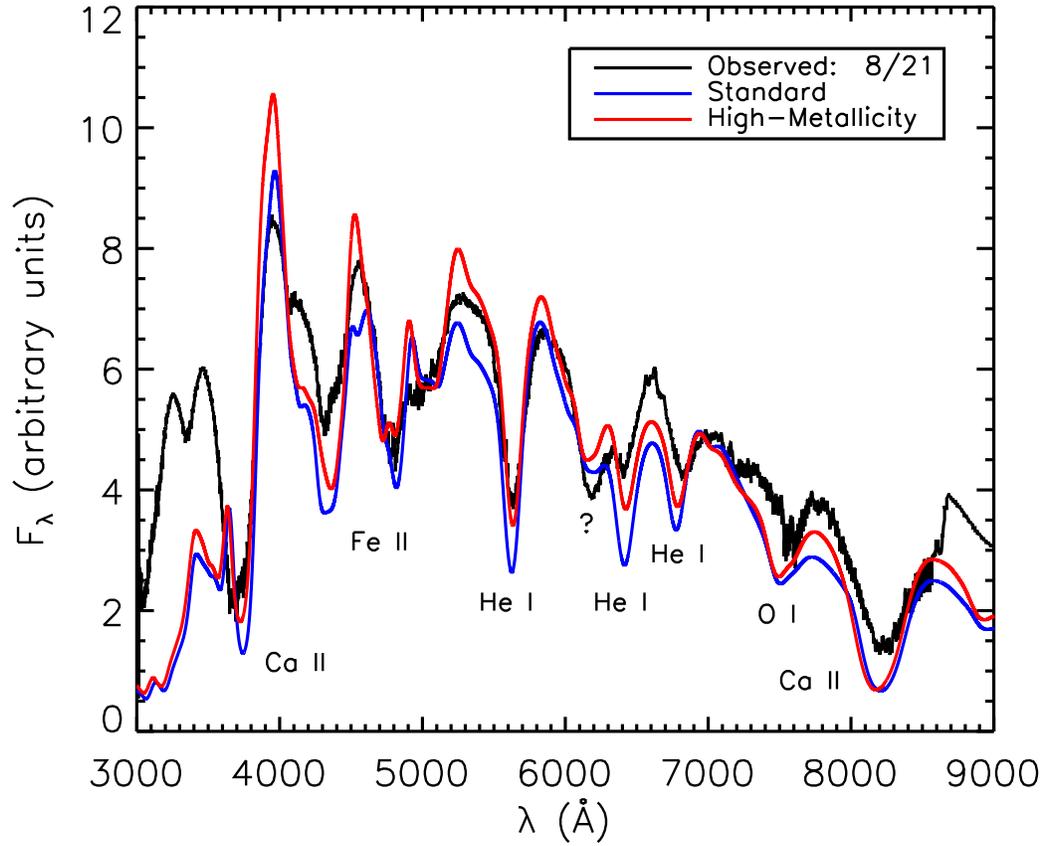}
\end{center}
\caption{\label{fig:aug21_both}The observed spectrum of SN~1999dn
  10 days before maximum light (obtained on August 21) is shown with
  two model spectra:  the standard composition model and a
  high-metallicity model.  Again, both models fit well, but the
  standard composition model fails to produce a strong feature near
  6200 \AA, while the high-metallicity model can.}
\end{figure}

Figure~\ref{fig:aug21_both} shows the observed spectrum obtained ten
days before maximum
light (epoch -10, obtained on August 21) with two spectral fits:  a
standard composition
model and a high-metallicity model.  The observed spectrum contains
similar features to those in the maximum light spectrum.  The two
models fit most of these features fairly well.  The Ca~II H and K
absorptions are slightly too strong in the models; likewise in the
emissions.  Both models fit the observed spectrum fairly well over the
region dominated by Fe~II lines, between 4100 \AA~and 5200 \AA.  The
He~I 5876 \AA~absorption is fit particularly
well by the high-metallicity model, while the emission is better
modeled with the standard composition.  Both models produce a feature
near 6200 \AA.  That feature is stronger and better fit to the observed
spectrum  in the high-metallicity model; however, it is hardly a
poor fit of the feature using the standard composition.  The He~I 6678
\AA~feature is particularly weak in the observed spectrum, a subtlety the
models fail to replicate.  Both models also appear to form the He I
7065 \AA~feature a little too far to the blue, which may be a
result of the simple $\gamma$-ray deposition that produces the
non-thermal excitation responsible for the formation of the neutral
helium lines in the models.  Likewise, the O~I absorption and the Ca~II
infrared triplet are formed a little blue in the model spectra, though
the features are otherwise fit reasonably well.

\begin{figure}
\leavevmode
\begin{center}
\includegraphics[width=0.75\textwidth,angle=90]{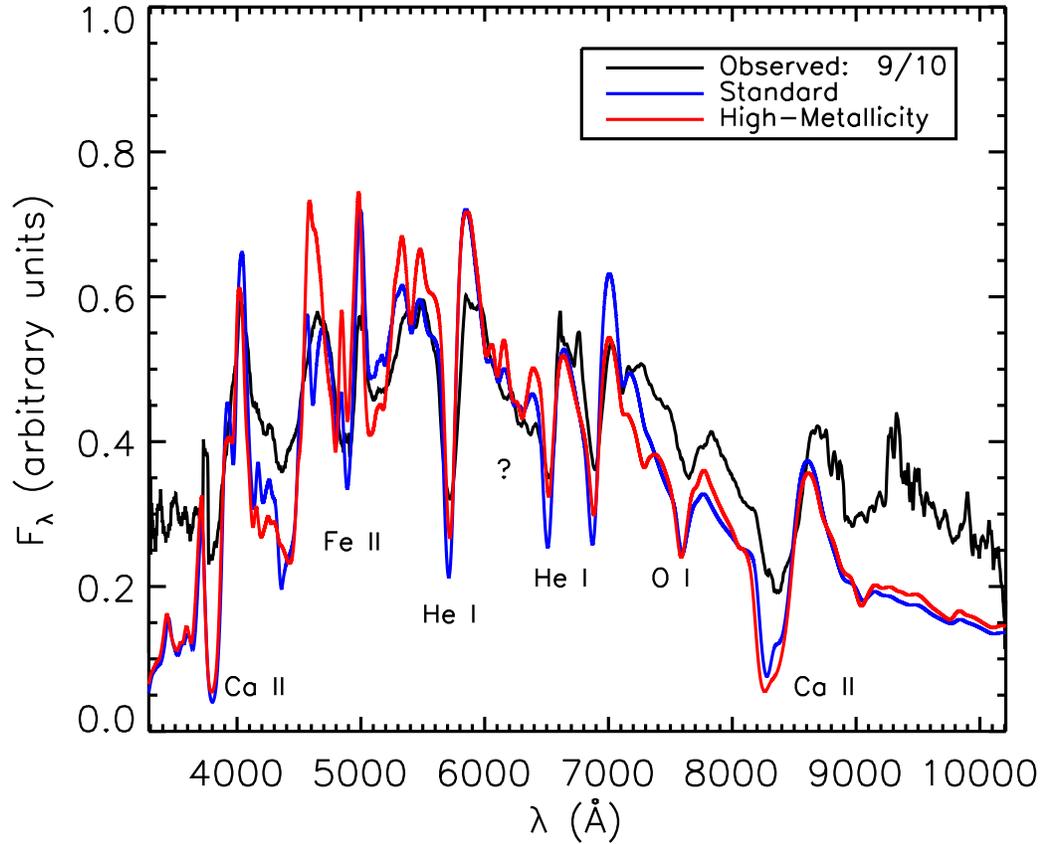}
\end{center}
\caption{\label{fig:sep10_both}The observed spectrum of SN~1999dn
  10 days after maximum light (obtained on September 10) is shown with
  two model spectra:  the standard composition model and a
  high-metallicity model.  While both models fit well, the standard
  composition model fits marginally better in the Fe II dominated
  region of the spectra between 4000 \AA~and 5000 \AA.  Also, the
  observed 6200 \AA~feature has become much weaker, with a better fit
  to it given by the standard composition model.}
\end{figure}

\begin{figure}
\leavevmode
\begin{center}
\includegraphics[width=0.75\textwidth,angle=90]{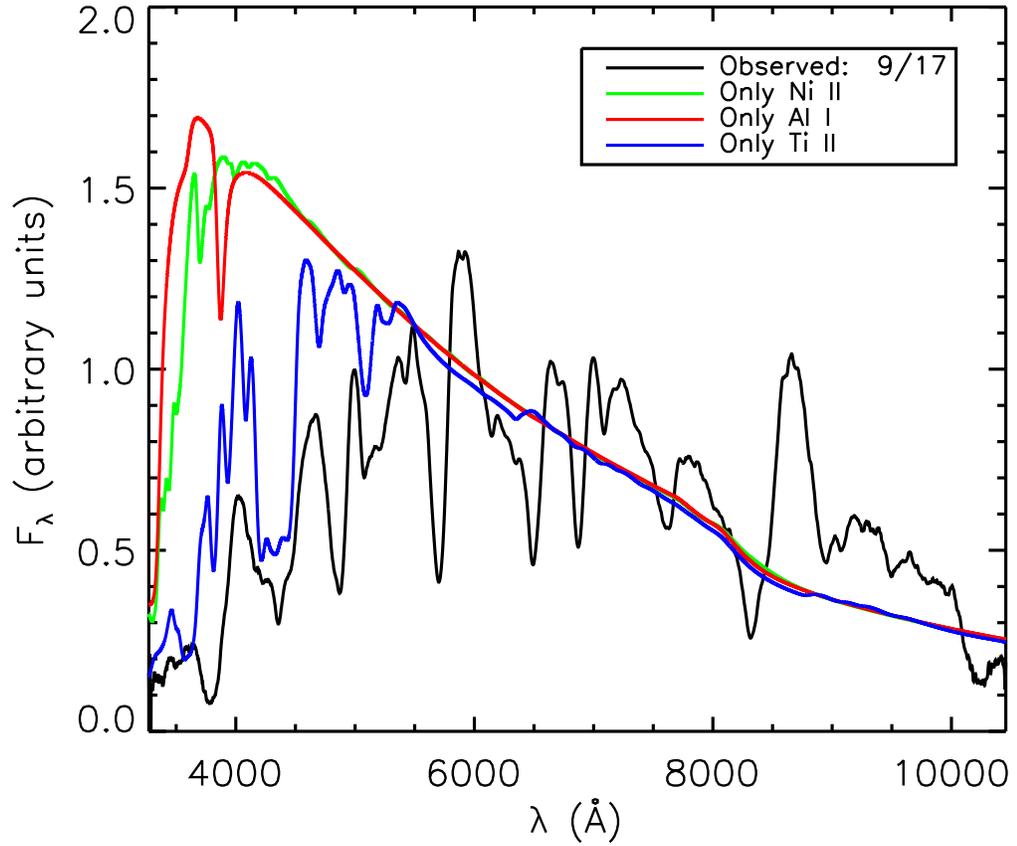}
\end{center}
\caption{\label{fig:sep17_LTE}The observed spectrum of SN~1999dn
  17 days after maximum light (obtained on September 17) is shown with
  single ion spectra of Ni II, Al I, and Ti II.   Note that Ti II has a
  very large contribution to the spectrum in the blue, in the Fe II region.}
\end{figure}

Moving to post-maximum light spectra, Figure~\ref{fig:sep10_both}
shows the observed spectrum
obtained ten days after maximum light (epoch +10, obtained September
10) with two spectral
fits:  again, a standard composition model and a high-metallicity
model.  The observed spectral features are largely the same as those
seen in the two previous epochs.  Both fits are of
good quality, though the standard
composition model fits the observed spectrum better overall,
particularly at shorter wavelengths.  Both
models have too strong an absorption for Ca~II H and K, but their
emissions fit well.  In the Fe~II dominated region (which is now
becoming blanketed by Ti II lines as well, see
Figure~\ref{fig:sep17_LTE}) the
high-metallicity model fails to fit as well as the standard
composition, which in turn struggles to fit some of the observed
features at the correct strengths.  (However, as opposed to the +0 day
spectrum, the feature at 5100 \AA~fits much better.)  The
standard composition model fits the He I 5876 \AA~feature more
closely, especially on the blue edge, but the high-metallicity model
fits the other two He~I absorptions marginally better.  The 6200
\AA~feature at this later epoch is not nearly as pronounced and has
evolved considerably from the feature at earlier times; the feature
actually appears to be two distinct features centered around 6200
\AA.  This is not unlike the proposed explanation for the
feature at earlier epochs, except now the two features are
resolvable.  In the region around 6200 \AA, the standard composition model reproduces
features that are more than strong enough to fit the observed.
Again, the He~I 7065
\AA, O I, and Ca II infrared triplet absorptions are produced a
little to the blue in the model spectra

\begin{figure}
\leavevmode
\begin{center}
\includegraphics[width=0.75\textwidth,angle=90]{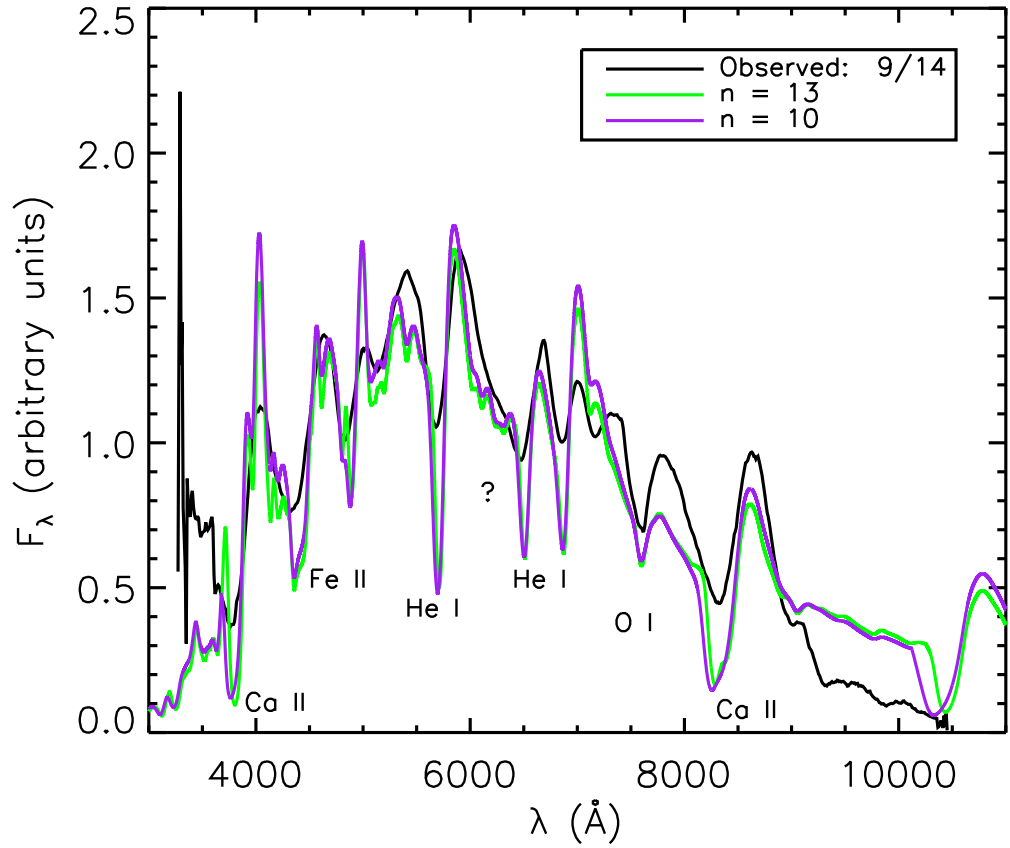}
\end{center}
\caption{\label{fig:sep14_n}The observed spectrum of SN~1999dn
  14 days after maximum light (obtained on September 14) is shown with
  the standard composition model while varying $n$ from 13 to 10.  At
  this epoch and in the +17 day models we prefer a less steep density
  profile, with the value $n = 10$.}
\end{figure}

In Figure~\ref{fig:sep14_n} the observed spectrum obtained 14 days
after maximum light (epoch +14, obtained September 14) is shown with
two spectral fits. Both models have standard composition, but they
vary between $n = 13$ and $n = 10$.  Both models fit the observed
spectra fairly well.  One main difference between the two are the line
strengths, particularly in the Fe~II dominated region.  The $n = 10$
model fits the strength of the absorptions better at 5300 \AA~and 4200
\AA.  The shapes of some of the absorption lines, like He I 5876
\AA~are also fit slightly better in the $n=10$ model.  Similar
observations can be made about later epochs as well.  Therefore, for
this and later epochs the models will have $n=10$, rather
than $n=13$.

\begin{figure}
\leavevmode
\begin{center}
\includegraphics[width=0.75\textwidth,angle=90]{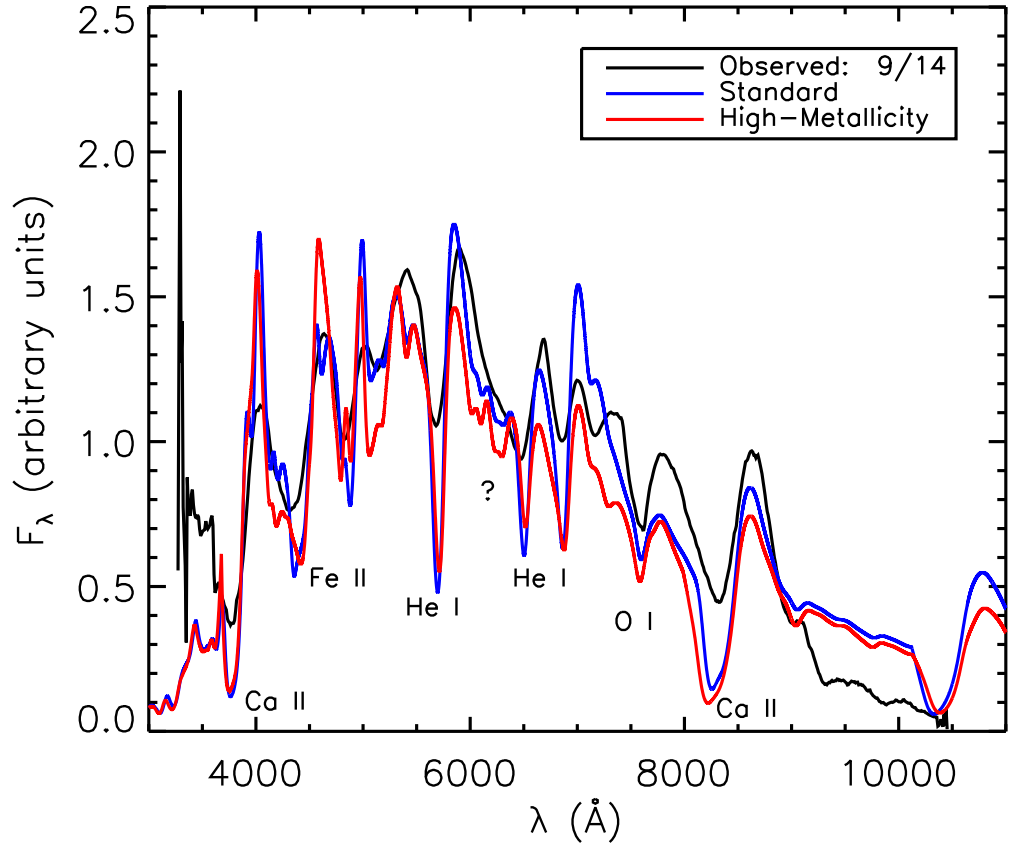}
\end{center}
\caption{\label{fig:sep14_both}The observed spectrum of SN~1999dn
  14 days after maximum light (obtained on September 14) is shown with
  two model spectra:  the standard composition model and a
  high-metallicity model, with $n = 10$ for both models.  Again, we
  prefer the standard composition model at this post-maximum epoch.}
\end{figure}

The September 14 observed spectrum is shown alongside two best fits
with varying metallicity in Figure~\ref{fig:sep14_both}:  one with the
standard composition and one with increased metallicity.  While both are good
fits overall, the standard composition model fits the observed
absorption depths better in the blue end of the spectrum, and
marginally worse in the redder end. Both models have absorptions and
emissions of the Ca II H and K lines that are stronger than
observed.  As mentioned above, the standard composition model fits the
absorption strengths in the Fe~II/Ti~II dominated region better than
the high-metallicity model.  Both models have strong absorption of He
I that fail to match the observed features perfectly. In the observed
spectrum barely anything remains of the 6200 \AA~feature; the models
produce features similar to those seen in the +10 day spectrum.  The
standard composition model fits the region much better because of the
decreased feature strength.  The models produce features that are at
the same wavelength of the observed spectrum for the He I 7065 \AA~and
O I lines; the Ca II infrared triplet absorption is only slightly too
blue, though a touch too strong, in the models compared to the
observed spectrum.

\begin{figure}
\leavevmode
\begin{center}
\includegraphics[width=0.75\textwidth,angle=90]{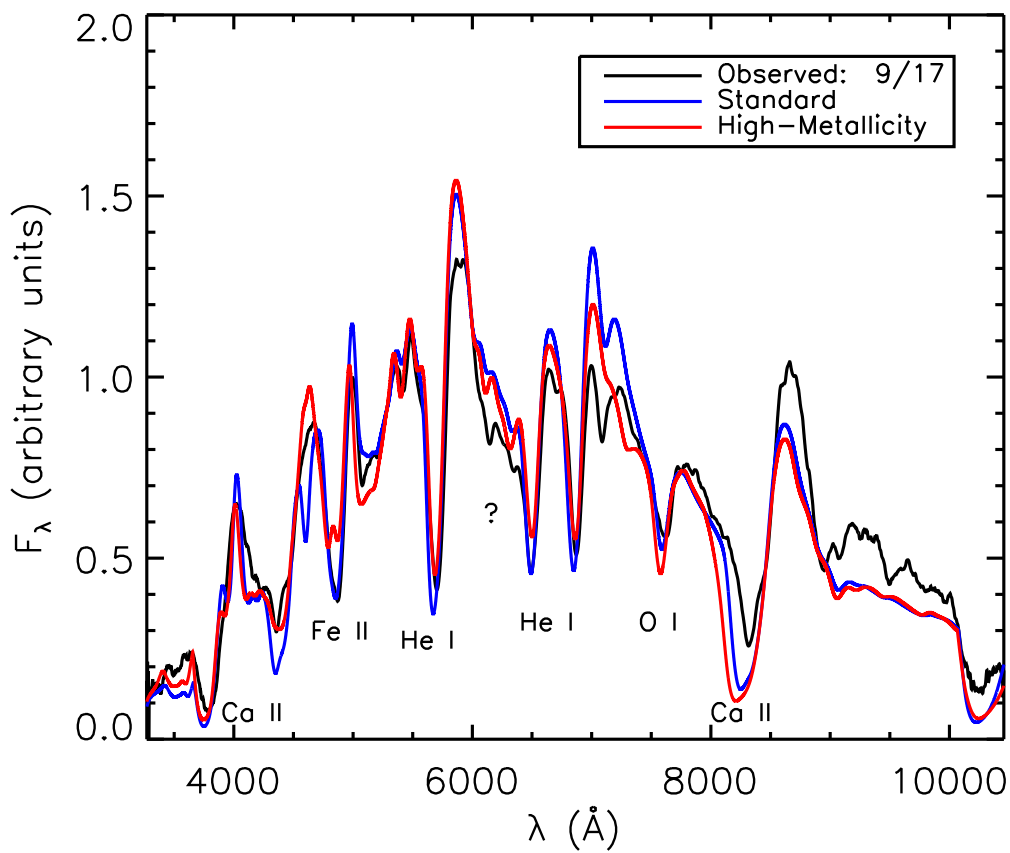}
\end{center}
\caption{\label{fig:sep17_both}The observed spectrum of SN~1999dn
  17 days after maximum light (obtained on September 17) is shown with
  two model spectra:  the standard composition model and a
  high-metallicity model.  Both models fit parts of the observed
  spectrum better than the other; overall, we prefer the standard
  composition model because of the decreased strength of the features
  in the blue.  The standard composition model fits the 6200
  \AA~feature slightly better as well.}
\end{figure}

Figure~\ref{fig:sep17_both} shows the observed spectrum of SN~1999dn
obtained 17 days after maximum light (epoch +17, obtained on September
17) with two spectral fits:  a standard composition model spectrum and a
high-metallicity model spectrum.  Both models fit the spectrum well.
The absorption and emission of the Ca~II H and K lines are fit
reasonably well in both models.  Neither model fits the Fe~II region
perfectly, but
both reproduce the features seen with varying degrees of accuracy:
the standard composition model fits the 4900 \AA~absorption perfectly,
while the high-metallicity model reproduces the features at 4400
\AA~and 5900 \AA~slightly better.  The standard composition model's
He~I 5876 \AA~absorption is slightly strong and blue, but the other
He~I absorption features are fit well.  Note that the features around
6200 \AA~are in the observed spectrum again, looking much like they
did in the +10 day spectrum. The shape of these features are reproduced
better in the standard composition model, though the absorption
strengths are reproduced better in the high-metallicity model.  The
features in the model spectra that
appear in the same region can be attributed to a blending of Fe~II,
Si~II, and C~II.  The O~I feature is fit well by both models (though
slightly better by the standard composition), and both models produce
a too strong and too blue Ca~II infrared triplet feature.

As the temperature in the supernova has decreased substantially since
maximum light (our model temperatures decrease from 6000 K to 5000
K), we expect to see some new contributions to spectral features at
later times.  Figure~\ref{fig:sep17_LTE} shows the observed spectrum
for the +17 day epoch along with three single-ion spectra of lines
that have only been considered in LTE:  Ni II, Al I, and Ti II.   Ni
II and Al I both have some effect on the model spectra close to the Ca
II H and K absorption (near 3750 \AA).  Ti II, however, has significant
contributions to the spectrum all the way up to a wavelength of 5500
\AA, in what we
previously called the Fe~II dominated region.  The presence
of Ti II lines was noted as a possibility in SYNOW analysis of this
spectrum by \citet{branIb02}.  The presence of Ti II in our models
is perhaps part of
the reason why the high-metallicity models fail to reproduce this
region of the spectrum as well as the standard composition models: 
they have too much Ti II, causing their absorptions to be too strong.
Ti II also has a very small feature in the 6200 \AA~region in our
models, which may affect the evolution of that
feature.

\begin{figure}
\leavevmode
\begin{center}
\includegraphics[width=0.75\textwidth,angle=90]{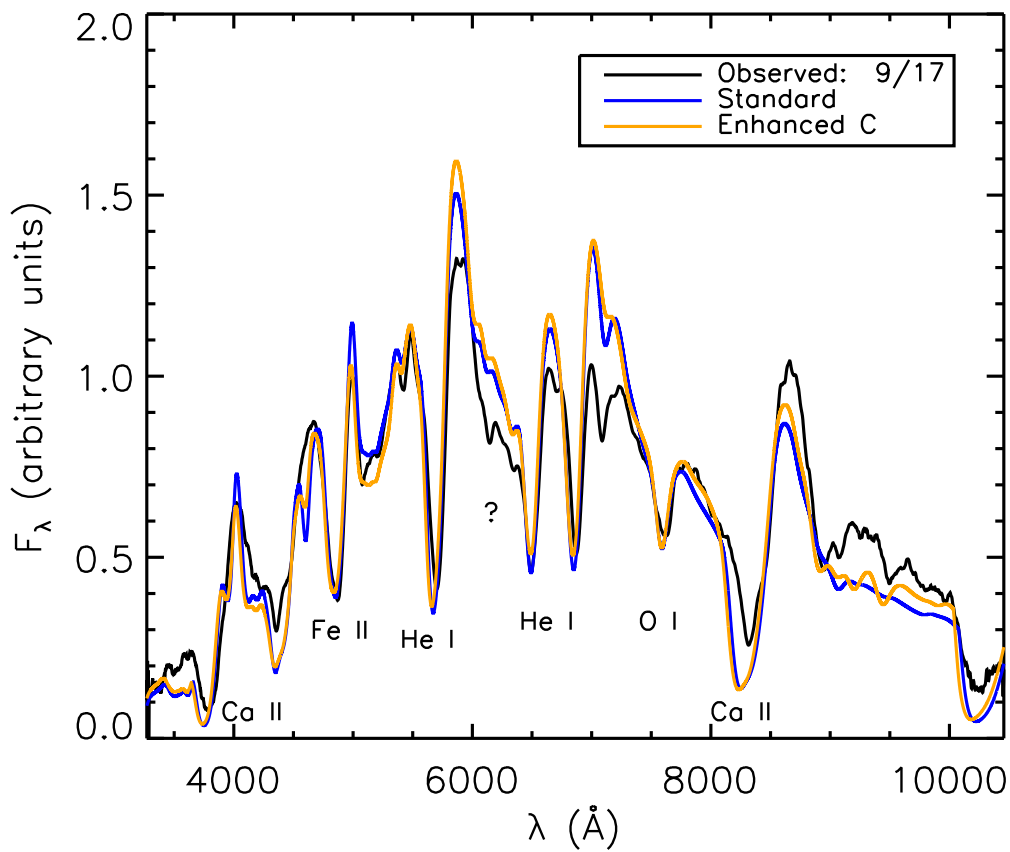}
\end{center}
\caption{\label{fig:sep17_10C}The observed spectrum of SN~1999dn
  17 days after maximum light (obtained on September 17) is shown with
  our standard composition model and a model with ten times as much
  carbon abundance.  Note that the enhanced carbon seems to have
  little effect in the 6200 \AA~region, but does reproduce an observed
  feature near 9500 \AA.  This may be evidence that there is increased
  C II at later times that may be changing the shape of some features
  in the spectrum, including the 6200 \AA~region.}
\end{figure}

Finally, we investigate the affects of increasing carbon abundance at
late times in Figure~\ref{fig:sep17_10C}.  \citet{deng00} claimed that
at +14 days C II was unambiguously in the spectrum.  Our late-time models show
some weak features of C II around the 6200 \AA~feature, but the effect
is minimal.  Increasing the carbon abundance by ten times in our +17
day standard composition model makes little effect on the overall
spectrum.  C II may be responsible for an observed feature near 9500
\AA~that is seen in the enhanced carbon model.  Unlike at earlier
epochs, near maximum light, C II has some affect on the shape of
particular features in the spectrum at late times, including around 6200
\AA, though its effect is small.

\section{Discussion}
Our goal in this series of papers is to (1) determine basic physical
properties of SN~1999dn at different epochs such as its effective
temperature and photospheric velocity, (2) determine if He I
absorption lines can be properly formed in a natural way by including
non-thermal excitation, and (3) determine whether there are spectral signatures
of hydrogen in this ``normal'' SN~Ib.  We have presented
models to fit five epochs of SN 1999dn that did not include hydrogen;
our models were homogeneous stellar atmospheres that had been
stripped of their hydrogen envelopes.  While this model is very
simple, our spectral fits are on the whole remarkably good by \phx\ standards.
Table~\ref{tab:models} contains a list of some of
the most basic physical parameters we determined, and in all epochs He
I absorptions were successfully produced in the models by including a
\gamray~deposition that followed the density of the ejecta that allowed for
non-thermal excitation.

Finally, we looked carefully at the prime
candidate for \halpha~absorption, a feature around 6200 \AA~in the
observed spectra, and tested alternative ways to reproduce the feature
in models.  The most successful, and a perfectly plausible,
alternative is an increased metallicity of the ejecta.  By increasing
the metallicity to three times solar, we were able to reproduce the
feature as a blend of Fe~II and Si~II lines well at early times
without affecting the rest of the spectrum too much.  At later times,
as the 6200 \AA~becomes weaker and the spectral contribution of metals
like Fe~II and Ti~II becomes more pronounced, the standard
composition models fit better, but we cannot rule out the
high-metallicity case.  We considered several other alternatives to the
identification of the 6200 \AA~feature:  Si~II alone created a feature
too far to the blue; C~II made no obvious effects in the standard
composition model at maximum light; and, Ne~I reproduced the feature,
but only after enhancing the neon abundance by a factor of 100.  Given
a uniform composition with \gamray~deposition following the density
profile of the ejecta, the most plausible alternative to
\halpha~appears to be a blending of Fe~II and Si~II, possibly
strengthened by increased metallicity.

We have noted that the 6200 \AA~feature evolves  with time.  At
early epochs it appears to be a single, strong  absorption feature, though this
single absorption could actually be a blend of more than one line, as
we showed in our high-metallicity models.  However, at later times,
the observed feature separates into multiple, distinct, weaker
features, possibly from  Fe~II, Si~II, C~II, and
Ti~II.  This change could be an indication that the 6200 \AA~feature
is \halpha~at early times, and is a blend of the above ions
at epochs after maximum light.  However, this evolution is also seen
in our models, suggesting that other physical properties play a role
in, and are perhaps important causes of the observed change.  Decreased
temperatures affect the shapes and strengths of the Fe~II and Si~II
lines, and allow other ions, like C~II and Ti~II, to become more
prominent in our models.  Decreased ejecta velocities cause less line
blending, allowing greater resolution of nearby features.
Furthermore, while our models do not include them as possibilities,
stratified compositions and density structures that do not follow a
single power law may affect the feature at later times as well.  With our
simple homogeneous parameterization, it is difficult to draw a strong
conclusion.  We hope to address the true nature and cause of this
evolution in Paper II.

In Paper II of this series, we will add a thin shell of hydrogen to
the outside of our current \phx\ models, and attempt to fine tune its velocity
coordinate and density profile to see if a model can reproduce the
6200 \AA~feature.

\acknowledgements
This work was supported in part by NASA grants NAG5-3505 and
NNG04GB36G, and NSF grant AST-0506028.  This research used resources
of the National Energy Research Scientific Computing Center (NERSC),
which is supported by the Office of Science of the U.S. Department of
Energy under Contract No. DE-AC03-76SF00098.  This research has
made use of the NASA/IPAC Extragalactic Database
(NED) which is operated by the Jet Propulsion Laboratory, California
Institute of Technology, under contract with the National Aeronautics
and Space Administration.

\clearpage

\clearpage
\begin{deluxetable}{ccc}
\tablecolumns{3}
\tablewidth{0pt}
\tablecaption{\label{tab:ions}  Ions Considered in NLTE}
\tablehead{}
\startdata
H I  & O I   & Si I\\
He I & O II  & Si II\\
He II& O III & Si III\\
C I  & Ne I  & Ca I\\
C II & Na I  & Ca II\\
C III& Na II & Ca III\\
N I  & Mg I  & Fe I\\
N II & Mg II & Fe II\\
N III& Mg III& Fe III\\
\enddata
\end{deluxetable}

\clearpage
\begin{deluxetable}{lccccc}
\tablecolumns{6}
\tablewidth{0pt}
\tablecaption{\label{tab:models} SN 1999dn Synthetic Spectra Model Parameters}
\tablehead{
  \colhead{Obs.~Spectrum Date} &
  \colhead{Epoch} &
  \colhead{Metallicity (\zsol)} &
  \colhead{\Tmod (K)} &
  \colhead{\vno (\kmps)} &
  \colhead{$n$}
}
\startdata
August 21\,\tablenotemark{a}    & -10 & 1 & 6750 & 11000 & 13\\
                                &     & 3 & 7250 & 11000 & 13\\
August 31\,\tablenotemark{a}    & +0  & 1 & 6000 & 10000 & 13\\
                                &     & 3 & 6500 & 10000 & 13\\
September 10\,\tablenotemark{b} & +10 & 1 & 5500 & 7000  & 13\\
                                &     & 3 & 6000 & 7000  & 13\\
September 14\,\tablenotemark{a} & +14 & 1 & 5500 & 7000  & 10\\
                                &     & 3 & 6000 & 7000  & 10\\
September 17\,\tablenotemark{b} & +17 & 1 & 5000 & 7000  & 10\\
                                &     & 3 & 5500 & 7000  & 10\\
\enddata

\tablenotetext{a}{From \citet{deng00}.}
\tablenotetext{b}{From \citet{math_sn1bc}.}

\end{deluxetable}

\clearpage

\clearpage

\clearpage

\clearpage

\clearpage

\clearpage

\clearpage

\clearpage

\clearpage

\clearpage

\clearpage

\clearpage

\end{document}